\documentclass[a4paper,12pt]{article}
\usepackage{latexsym}
\usepackage{amsmath}
\usepackage{amsfonts}
\usepackage{amssymb}
\usepackage{graphicx}
\usepackage[latin1]{inputenc}
\usepackage{color}
\begin{document}

\title{Exactly solvable Gaussian and  non-Gaussian  mean-field games and collective swarms dynamics }
\author{\\ \\ Max-Olivier Hongler\thanks{max.hongler@epfl.ch}\\  \\ Ecole Polytechnique F\'ed\'erale de Lausanne (EPFL) \\ Sciences et Techniques de l'Ing\'enieur (STI)}
\date{}
\maketitle

\abstract{ 

\noindent The collective behaviour  of stochastic  multi-agents swarms   driven by  Gaussian and non-Gaussian environments is  analytically discussed in  a mean-field approach.  We first exogenously implement long range mutual interactions  rules with strengths that are modulated by  the real-time distance  separating  each agent  with  the swarm barycentre. Depending on the form of this barycentric modulation, a  transition  between  drastically collective behaviours can be unveiled. A behavioural  bifurcation threshold  due to  the tradeoff between the desynchronisation effects of  the  stochastic environment   and the  synchronising interactions is analytically  calculated.  For strong enough  interactions, the emergence of a swarm  soliton  propagating wave is observable. Alternatively, weaker interactions cannot overcome the environmental noise and  evanescent diffusive waves result.  In a second and complementary  approach, we show the the  emergent solitons  can alternatively be interpreted as being the optimal equilibrium   of mean-field games (MFG)  models
 with  ad-hoc running cost functions which are  here exactly  determined. The MFG's  equilibria  resulting  from the optimisation  of individual utility functions  are solitons  that are  therefore endogenously  generated.  Hence for the classes of models here proposed, an explicit  correspondence between exogenous and endogenous  interaction rules ultimately  producing similar collective effects can be explicitly constructed.   For both Gaussian and non-Gaussian  environments our exact results unveil new classes of exactly solvable mean-field games dynamics. }

\vspace{0.5cm}


%

\section{Introduction}\label{prifon}

\noindent  Yoshiki  Kuramoto proposed for the first  time in 1975  a fully analytic study  describing   the collective behaviour of a swarm of interacting Brownian  phase oscillators  [1].  In this now paradigmatic multi-agent model,  each phase oscillator evolving  on a one-dimensional  compact state space ({\it circle})  interacts with all its neighbours ({\it long-range type  interactions}). The dynamics is described by a set of coupled  stochastic differential equations (SDE)  driven by  independent White Gaussian Noise (WGN).  In the very  large population limit,  it is legitimate to use of  a mean-field (MF) approach enabling to summarise the swarm behaviour into an {\it oscillator density measure}. Thanks to the underlying WGN the density measure evolves  according to a nonlinear deterministic Fokker-Planck equation (FPE) which, in the stationary regime,  can be analytically discussed. The trade-off between the desynchronising  tendency due to the random environment and  the synchronising effect due to the mutual interaction leads to bifurcation threshold separating two drastically different swarm behaviours. Namely for low coupling to noise ratio, the collective motion is  diffusive and fully disorganised, for large coupling to noise ratio however, a collective synchronised  swarm emerges. While for  the  basic Kuramoto model, the mutual interactions rule leading to the collective behaviour  is exogenously given,  a recent  approach [2] shows how the Kuramoto's  corporative behaviour  may alternatively be  viewed as being an equilibrium of a mean-field games (MFG) formalism pioneered in [3, 4, 5, 6, 7]  with ad-hoc running cost functions. In this MFG approach each agent (i.e. phase oscillator) minimises an individual cost function which depends on the whole swarm and it is this collective minimisation procedure which  leads to a global equilibrium which coincides with  the Kuramoto's  synchronised  phase.  Adopting the alternative MFG  point of view,  one may interpret the collective behaviour as being the result of an endogenous rule (each agent possesses its  own objective function to minimise).  We hope that our present models  offer an additional  elementary explicitly solvable illustration of the  recent mathematically oriented   literature  devoted to MFG dynamics [8, 9, 10, 11,12].

\vspace{0.3cm}
\noindent The central  goal in this paper  is to construct fully solvable  classes of scalar  multi-agents models with state space $\mathbb{R}$ instead of the compact Kuramoto's  circular state space.  The underlying stochastic environment  is either the Brownian motion process or a two-states Markov chain in  continuous time (also known as the {\it telegraphic noise}) which introduces correlations into the dynamics.  To separately discuss the resulting dynamics, our presentation proceeds via two main sections.  In section \ref{difon} the  agents' dynamics  are driven  by Brownian environment sources and Propostion 1 shows how a soliton propagating wave emerges from an exogenously specified  algorithm referred as the {\it avoid to be the laggard}  (ABL) rule. Depending on the strength  of a barycentric  factor which modulates the actual  influence of  agents  depending of their  locations relative  to the swarm's barycentre,  a behavioural transition,  of the Kuramoto's type arises and the exact bifurcation threshold can be  calculated. In Proposition 2, we construct  associated MFG's  with equilibria that are determined by solving a nonlinear  Schr\"odinger equation.  This yields  soliton waves similar to those  obtained in Proposition 1.  A similar  presentation architecture  adopted  in section \ref{trifon}  where the  corresponding Propositions 3 and 4  are obtained in presence of  telegraphic noise environments.  Proposition 4,  unveils  a  new class of  analytically solvable MFG (i.e. not  belonging to the linear drift with quadratic costs  optimal control dynamics).


\section{Nonlinear diffusive dynamics}\label{difon}

 \noindent Let us consider a set of $N$ scalar  interacting diffusion processes $X_{k,t} \in \mathbb{R}$ with time  $ t \in \mathbb{R}^{+}$ :
 
 \begin{equation}
\label{BASE}
dX_{k,t} = {\cal I}({\bf X}_t, X_{k,t}) dt + \sigma dB_{k,t}, \qquad \qquad  k=1,2,..,N,
\end{equation}
 
 \noindent with ${\bf X}_t : =( X_{1,t}, X_{2,t}, \cdots, X_{N,t})$, $\sigma \in \mathbb{R}^{+}$ and $ dB_{k,t}$ are $N$ independent standard Brownian motions [13]. The drift ${\cal I}({\bf X}_t, X_{k,t}) $ defines a mutual-interaction kernel exogenously implementing the  algorithm:

 \vspace{0.3cm}
 \noindent{\bf Avoid being a laggard algorithm (ABL) }.

 \begin{itemize}
  \item[] i)    For $k=1,2,\cdots,N$ and in real time,  agent ${\cal A}_k$   observes the positions $X_{j,t}$ of his fellows  ${\cal A}_j$ for $j \neq k$ and $j=1,2,\cdots,N$.

  \item[] ii)   For $k=1,2,\cdots ,N$ agent ${\cal A}_k$  accounts the number  $n_k(t) $ of  her leaders ${\cal A}_{j}$ for which  $X_{j,t} \geq X_{k,t}$ and  for $j\neq k$.   
  
    \item[] iii)  For $k=1,2,\cdots ,N$  agent ${\cal A}_k$  implements her instantaneous drift  according to the rule:
    
    \begin{equation}
\label{RULE}
{\cal I}({\bf X}_t, X_{k,t}) ={ n_k(t) \over N}.
\end{equation}
    
\noindent In view of Eq.(\ref{RULE}),  ${\cal A}_k$   effectively avoids to remain a swarm's laggard since the more leaders she finds, the higher is her incentive to increase a drift velocity.  
    
 \end{itemize}

\noindent    
 
\noindent In the sequel, we will systematically focus attention on  large populations (i.e. $N \rightarrow \infty$) enabling us define a empirical agents population  density $\rho(x,t) \in [0,1]$ as 

\begin{equation}
\label{EMPIRIC}
\rho(x,t) = \displaystyle {1 \over N}\sum_{j=1}^{N} \delta ( X_{j,t} -x) .
\end{equation}

\noindent Since the agents  population is homogeneous, (i.e. ${\cal I}_{k}(\cdot) \equiv {\cal I}(\cdot)$), we may randomly select one representative  (i.e. index independent)  fellow ${\cal A}$ located at  $X_t \in \mathbb{R}$. For ${\cal A}$, the ABL rule is formally implemented as:   
\begin{equation}
\label{SDEBASE}
dX_t =  \left[ \int_{X_T}^{\infty} \rho(y,t) dy \right] dt + \sigma dB_t.
\end{equation}

\noindent {\bf Remark 1} ({\it mean-field dynamics}).  Eq.(\ref{SDEBASE})    implements   {\it  infinite  range}  interactions since agent ${\cal A}$ has to take into account  the locations of the whole  swarm population  (except herself)  to determine her own drift. This basically realises  the {\it mean-field }approach  of the swarm's dynamics.

\vspace{0.3cm}
\noindent The probabilistic properties of the trajectories solving the (Markovian) stochastic differential equation (SDE)   Eq.(\ref{SDEBASE})   can be  found by solving the associated nonlinear Fokker-Planck equation (FPE) [13]:

\begin{equation}
\label{DIFFNLO}
\partial_t \rho(x,t) = {\sigma^{2}\over 2} \partial_{xx} \rho(x,t)   - \partial_x\left(\rho(x,t) \left[\int_x^{\infty} \rho(y,t) dy\right] \right).
\end{equation}

\noindent Let us now    generalise the ABL rule Eq.(\ref{SDEBASE})   by further introducing an  (infinitely  differentiable)  {\it barycentric weighting function}   ${\cal G}\left[ X_t- \langle X(t)\rangle\right] : \mathbb{R}  \rightarrow \mathbb{R}^{+}$. Accordingly,  Eq.(\ref{DIFFNLO})  will be now generalised as:
\begin{equation}
\label{DIFFNL}
\left\{
\begin{array}{l}
\partial_t \rho(x,t) = {\sigma^{2}\over 2} \partial_{xx} \rho(x,t)   - \partial_x\left\{\rho(x,t) \int_x^{\infty} \ \left({\cal G} \left[ y- \langle X(t)\rangle \right] \rho(y,t) dy\right) \right\},\\ \\

\langle X(t) \rangle = \int_{\mathbb{R}} x \rho(x,t) dx, \\ \\
 \int_{\mathbb{R}} \rho(x,t) dx =1,
\end{array}
\right.
\end{equation}

\noindent  For agent ${\cal A}$, the weight  ${\cal G}(\cdot) $   modulates  the relative influence of the leaders depending on their relative remoteness to  the swarm barycentre. For $z$-increasing ${\cal G}(z)$ we effectively describe situations where leaders are  more influential than close neighbours. Conversely   decreasing ${\cal G}(z)$  describe dynamics where agents  are mostly influenced  by their neighbouring fellows. It is important to point out  that in both situations  the interactions remain  of long range type. For the nonlinear swarm dynamics expressed in Eq.(\ref{DIFFNL}), we now can establish:

\vspace{0.5cm}
\noindent {\bf Proposition 1}. 

\noindent {\it Assuming  the class of kernel functions  $ {\cal G}_{\eta , \sigma}(x) :=A(\eta,  \sigma) \cosh(x)^{\eta}$  with parameters $ \sigma\in \mathbb{R}^{+}$,  $\eta \in ]-2,\infty]$ and  the pre-factor $A(\eta, \sigma)= {\sigma^{2} (2+ \eta)\over 2   {\cal N}(\eta)}$, Eq.(\ref{DIFFNL})  is solved by the normalised  soliton   propagating waves:

\begin{equation}
\label{SOLVAR}
\left\{
\begin{array}{l}
\rho(x, t) = {\cal N} (\eta)\cosh^{-(2+ \eta)}( x- \omega t), \qquad (2+ \eta) >0,\\ \\
 \omega =   {\cal N} (\eta) A ( \eta, \sigma), \\ \\
 {\cal N} (\eta)^{-1}= B(1/2, 1 + \eta/2) ={ \sqrt{\pi} \Gamma( 1 + \eta/2) \over  \Gamma(3/2+\eta/2)},
\end{array}
\right.
\end{equation}}

\noindent where $B(1/2, 1 + \eta/2) :={ \sqrt{\pi} \Gamma( 1 + \eta/2) \over  \Gamma(3/2+\eta/2)}$ is the beta function [14].

\vspace{0.5cm}
\noindent {\bf Proof of proposition 1}. 

\noindent Introduce the change of variables $x \mapsto \xi = (x- \omega t)$ and $t \mapsto \tau$ and assume   the $\tau$-independent  stationary evolution $ \rho(\xi, \tau) =  \rho(\xi) = {\cal N}(\eta) \cosh^{-(2+\eta)}(\xi)$. Accordingly, we have:

$$
\left\{
\begin{array}{l}
\langle X(t) \rangle = \int_{\mathbb{R}} x \rho(x,t) dx=\int_{\mathbb{R} }( \xi + \omega t) \rho(\xi) d\xi  =  \omega t, \\ \\ 

\partial_t  \mapsto -\omega \partial_{\xi} + \partial_{\tau} \qquad {\rm and} \qquad   \partial_x \mapsto \partial_{\xi}, \\ \\
\int_x^{\infty} \ \left[{\cal G}_{\eta, \sigma}( y- \langle X(t)\rangle ) \rho(y,t) dy\right] \mapsto \int_{\xi}^{\infty} {\cal G }_{\eta, \sigma} (\xi) \rho(\xi) d\xi = \int_{\xi}^{\infty} A( \sigma, \eta) {\cal N} (\eta) \cosh^{-2} (\xi) d\xi .
\end{array}
\right.
$$

\noindent It follows that Eq.(\ref{DIFFNL})  can be rewritten as:

$$
\partial_ {\xi} \left\{ {\sigma^{2}\over 2}  \partial_{\xi} \rho(\xi)  + \omega  \rho(\xi) - \rho(\xi) \int_{\xi}^{\infty} A( \sigma, \eta)  {\cal N} (\eta) \cosh^{-2} (\xi) d\xi  \right\} =0.
$$

\noindent Integrating once the last equation with respect to $\xi$ (with vanishing integration constant in order  to fulfil the normalisation constraint,  one has to impose $\lim_{|\xi| \rightarrow \infty} \rho(\xi)  =0$) and dividing by  $\rho(\xi) >0$, we straightforwardly  obtain:

$$
 {\sigma^{2}\over 2}  \partial_{\xi} \log [\rho(\xi)]  =  - \omega + \int_{\xi}^{\infty} A( \sigma, \eta) {\cal N} (\eta) \cosh^{-2} (\xi) d\xi.  
$$

\noindent Plugging   $\rho(\xi) = {\cal N} (\eta) \cosh^{-(2+\eta)} (\xi)$  into the last equation and using  $\int_{\xi}^{\infty} \cosh^{-2} (\xi )  d\xi = 1- \tanh(\xi)$, the last  expression reads:

$$
- {\sigma^{2}\over 2}  (2+\eta) \tanh(\xi)  =   - \omega +  A(\sigma, \eta)    {\cal N} (\eta)  [1-\tanh(\xi)].
$$

\noindent By direct identification, we see that we  need to fulfil:

\begin{equation}
\label{DETER}
\left\{
\begin{array}{l}

   A( \sigma,\eta)= {\sigma^{2}  (2+\eta) \over 2 {\cal N}(\eta)}, \\ \\ 
\omega =   A (\sigma, \eta)  {\cal N} (\eta).

\end{array}
\right.
\end{equation}

\noindent The normalisation  factor ${\cal N} (\eta)$ imposes  $(2 + \eta) >0$  and, invoking  [14], we have:

$$
{\cal N} (\eta)^{-1} = \int_{\mathbb{R}} \cosh^{-(2+ \eta)}(x)dx = B\left({1 \over 2} , 1 + {\eta \over 2} \right).
$$

 \rightline{{\bf End of the proof}}

\vspace{0.5cm}
\noindent {\bf Remark 1}. It is worth to observe that a normalised soliton cannot be generated  for weights ${\cal G}(z) \propto\cosh^{\eta} (z)$  for $\eta < -2$. This can be heuristically understood by  the fact that decreasing   $\eta$, reduces  the cooperative influence of  remote leaders. This  weakens the possibility to sustain a collective evolution. Hence $\eta =2$ is the bifurcation value separating two drastically different swarm evolution, namely for $\eta > -2$, one observes the emergence of a  co-operative  soliton which cannot be sustained when $\eta \leq -2$.


\subsection{Corresponding mean-field game dynamics}
\noindent Consider the  diffusive dynamics:

\begin{equation}
\label{DRIVE}
dX_t = \left[ b + u(X_t) \right]dt + {\sigma^{2} \over 2} dB_t,
\end{equation}

\noindent with $b$ a constant and $u(X_t)$ a control function dependent on the whole population of agents. In parallel, we now introduce a  MF  running cost function  ${\cal L}[u, \rho(x,t) ]$ in the form [14]:

\begin{equation}
\label{CF}
 {\cal L}[u, \rho(x,t) ] := {1 \over 2 \mu}\left[ u(x) \right]^{2} - g \left[\rho(x,t) \right]^{a}.
\end{equation}

\noindent with $g, \mu , a \in \mathbb{R}^{+}$ For a time horizon $T$, we introduce for $t \in [0,T]$  a cost functional:

\begin{equation}
\label{STOPT}
{\cal J} [X(\cdot), u(\cdot)] = \mathbb{E} \left\{  \int_{0}^{T}{\cal L}[u(X_s), \rho(X_s,s)] ds\right\} + C_{T}(X_T).  
\end{equation}

\noindent where  $C_{T}(X_T)$ stands for a  final cost.  Minimisation of the cost given by Eq.(\ref{STOPT})  leads  to a set of nonlinear coupled  pde's  which have to be simultaneously solved forward/backward in time [15]:

\begin{equation}
\label{FORBACK}
\left\{
\begin{array}{l}
\partial_t \rho(x,t) = \partial_{x} \left[ \left(  {1 \over \mu}\partial_x    u(x,t) 
  - b \right) \rho(x,t) \right] 
+ {\sigma^{2} \over 2} 
\partial_{xx}  \rho(x,t), \qquad\qquad \qquad \, {\rm (Fokker\,\,Planck),}\\ \\ 
\partial_t u(x,t) +  b \partial_x u(x,t)- {1 \over 2\mu} \left[ \partial_x u(x,t) \right]^{2} +  {\sigma^{2} \over 2} 
\partial_{xx}  u(x,t) = 
g \left[\rho(x,t) \right]^{a}, \, {\rm (Hamilton\,\,Bellman\,\,Jacobi).}
\end{array}
\right.
\end{equation}

\noindent Assume now  that we deal with sufficiently large time horizons $T$  so that for the range of times $0 << t << T$, an ergodic  regime [12, 15, 16] can be reached. In this stationary regime, we approximately have:

\begin{equation}
\label{ERGODIC}
{u(x,t) \over t}  \quad \displaystyle \simeq \epsilon \quad {\rm for } \,\,\,  0<<t<<T,
\end{equation}

\noindent with $\epsilon$ an a priori unknown   constant. In this time range, the initial conditions and  the final cost barely  affect the solution of  Eq.(\ref{FORBACK})  and we can establish:

\vspace{0.5cm}
\noindent {\bf Proposition 2}.

\noindent {\it Given the parameters $b$, $\sigma$ in Eq.(\ref{DRIVE}) and $a$, $\mu$  in Eq.(\ref{CF}) and for   $g = {\mu \sigma^{4} (a+1) \left[ B \left({1 \over 2} , {1\over  2a} \right)\right]^{2a} \over 2 a^{2}  }$ in Eq.(\ref{CF}),  the probability density $\rho(x,t)$ associated with ergodic regime of  the MFG dynamics  defined by Eq.(\ref{FORBACK})  reads:

\begin{equation}
\label{MFGSOLU1}
\rho(x,t)  = {B \left({1 \over 2} , {1\over  2a} \right) \over \left[\cosh(x- b t) \right]^{1/a}}, \\ \\ 
\end{equation}}


\vspace{1cm}
\noindent {\bf Proof of Proposition 2}

\noindent We first  introduce a couple of  auxiliary scalar fields $\left[\Phi(x,t), \Psi(x,t)\right]$ defined by:

\begin{equation}
\label{SWIE}
\left\{
\begin{array}{l}
\Phi(x,t) =  e^{ - \left[ { u(x,t) - \epsilon t  \over \mu \sigma^{2}} \right]}, \\ \\
\Psi(x,t)  =    e^{ + \left[{ u(x,t) - \epsilon t\over \mu \sigma^{2}}\right]} \rho(x,t) 
\end{array}
\right.
\end{equation}

\noindent and hence $\rho(x,t) = \Phi(x,t) \Psi(x,t)$.  In terms of   $\left[\Phi(x,t), \Psi(x,t)\right]$, Eq.(\ref{FORBACK}) can be rewritten as, (see Appendix):

\begin{equation}
\label{FORBACK2}
\left\{
\begin{array}{l}
 \epsilon \Phi(x,t) -\mu \sigma^{2} \partial_t \Phi(x,t) =  + \mu \sigma^{2} b \partial_{x}\Phi(x, \tau) + {\mu \sigma^{4} \over 2} \partial_{xx}  \Phi(x,t)  + g \left[ \Phi(x,t) \Psi(x,t) \right]^{a} \Phi(x,t) , \\ \\ 

\epsilon \Psi(x,t)+ \mu \sigma^{2} \partial_t \Psi(x,t) =    - \mu \sigma^{2} b \partial_{x}\Psi(x, \tau)+{\mu \sigma^{4} \over 2} \partial_{xx}  \Psi(y,t)  + g \left[ \Phi(x,t) \Psi(x,t) \right]^{a} \Psi(x,t).
\end{array}
\right.
\end{equation}

\noindent Introduce the Galilean frame of coordinates $(\tau, \xi)$ defined by:

\begin{equation}
\label{GAL}
t \mapsto \tau , \quad x \mapsto \xi = x- bt \qquad \Rightarrow \qquad  \partial_t \mapsto \partial_{\tau}  - b \partial_{\xi}, \quad \partial_x \mapsto \partial_{\xi}
\end{equation}

\noindent  implying that  Eq.(\ref{FORBACK2}) takes the form: 

\begin{equation}
\label{NLSCHR}
\left\{
\begin{array}{l}
 \epsilon \Phi(\xi, \tau) -\mu \sigma^{2} \partial_{\tau}\Phi(\xi, \tau) =  {\mu \sigma^{4} \over 2} \partial_{\xi \xi}  \Phi(\xi, \tau )  + g \left[ \Phi(\xi, \tau) \Psi(\xi, \tau) \right]^{a} \Phi(\xi, \tau), \\ \\
 \epsilon \Psi(\xi, \tau)+ \mu \sigma^{2} \partial_\tau \Psi(\xi, \tau)   =    {\mu \sigma^{4} \over 2} \partial_{\xi \xi}  \Psi(\xi, \tau ) + g \left[ \Phi(\xi, \tau )\Psi(\xi, \tau \right)]^{a} \Psi(\xi, \tau).
 
 \end{array}
 \right.
\end{equation}

\noindent In the  stationary regime  reached when $\partial_{\tau}\Phi(\xi, \tau)= \partial_{\tau}\Psi(\xi, \tau)=0$, the resulting nonlinear ODE's  for $\Psi(\xi)$ and $\Phi(\xi)$  coincide and are formally similar to a nonlinear Schr\"odinger equation [15]. Imposing therefore  that $\Psi(\xi)=\Phi(\xi)$  and  integrating  the $\tau$-independent version of Eq.(\ref{NLSCHR}) by separation of variables, we obtain:

\begin{equation}
\label{I1}
\int { d \Phi (\xi) \over  \sqrt{A_1 (\epsilon)\Phi^{2}(\xi) - A_2(g)\Phi^{2a+2} (\xi)} }= \xi, \qquad A_1(\epsilon) ={ 2\epsilon \over \mu \sigma^{4}} \quad {\rm and}\quad A_2(g)= {2g \over (a+1) \mu \sigma^{4} }.
\end{equation}

\noindent Using the identity $\cosh^{2}(x) -1= \sinh^{2}(x)$, it s straightforward to verify   that 

 \noindent provided we have:
$$
\left\{
\begin{array}{l}
A_1(\epsilon) a^{2} =1  \quad \Rightarrow \quad \epsilon ={1 \over 2}  \mu \sigma^{4} a^{2} ,\\ \\

 A_2(g) a^{2}  {\cal N}(a)^{2a}  =1, \quad \Rightarrow \quad   g = {\mu \sigma^{4} (a+1) \left[ {\cal N}(a) \right]^{-2a} \over 2 a^{2}  },
\end{array}
\right.
$$

 \noindent the normalised  solution of Eq.(\ref{I1}) reads as :

\begin{equation}
\label{SOLIDE}
\Phi(\xi) = \Phi (x- bt) = {{\cal N} (a)\over \cosh^{1/a}( x- bt)}, 
\end{equation}

\noindent  with  the normalisation factor ${\cal N}(a)$ is given by [13] :

$$
{\cal N}(a)^{-1}=  \int_{\mathbb{R}} {d\xi  \over \left[ \cosh(\xi) \right]^{1/a}}   = B(1/2 , 1/2a).
$$

\rightline{{\bf End of the proof}}

\vspace{1cm}
\noindent  {\bf Remark 2}. By  identifying   $ \omega = b$ and $a^{-1} = (2 + \eta)>0$,  we see that both  solitons arising in  Eqs.(\ref{SOLVAR}) and (\ref{SOLIDE})  are identical. Therefore, one can directly assert that  the  exogenously given ABL algorithm generates a collective behaviour which coincides with the  optimal ergodic equilibrium of the MFG with running costs function  ${\cal L}[u, \rho(x,t) ]$ given by Eq.(\ref{CF}). Large $a$ parameters in Eq.(\ref{CF}) describe MFG situations where interaction costs are confined to close  neighbours and hence relatively small  corresponding parameters  $\eta$. Conversely small $a$'s lead to MFG with widely spread interactions  to which corresponds large $\eta$'s in the  exogenously defined ABL rule.


\section{Piecewise evolution dynamics}\label{trifon}

\noindent  In this section, the  exposition delivered   in section \ref{difon} will be  repeated for environments driven by two states Markov chains in continuous time (i.e.  {\it telegraphic process}) [17]. Instead of the diffusion dynamics given by Eq.(\ref{BASE}), we shall here consider a  set of discrete two velocity Boltzman's equation, (i.e.  also known as the Ruijgrok-Wu dynamics (RW) [18]) with random Poisson switchings between the two velocity states. Specifically, one considers a set of $N$ agents evolving on $\mathbb{R}$ evolving with velocities either $-1$ or $+1$.  Driven by a couple of  Poisson processes with rates $u_{\pm}\geq 0$,  the  agents' velocities spontaneously switch from  $-1$ to $+1$ (respectively  $+1$ to $-1$) velocity states. In addition to the spontaneous switchings, a Boltzman nonlinear  collision term  implies that  when a pair of  particules with velocities $(-1, +1)$  collide, it emerges with a given  rate  a $(+1,+1)$  pair. In the sequel, we shall  assume that the $(-1, +1)$  collision rate is itself modulated by the instantaneous  configuration of the swarm of particles. For  $x \in \mathbb{R}$ and time  $t\in \mathbb{R}^{+}$,  $P(x,t)dx $ (respectively $Q(x,t)dx $)  stand for  the proportion  of agents with velocities $+1$   (respectively $-1$)  located in $[x, x+dx]$.  Instead of the Fokker-Planck Eq.(\ref{DIFFNL}), the corresponding  dynamics now reads  as a generalised version of the velocity Boltzman's equation:

 {\footnotesize \begin{equation}
\label{TELEG}
\left\{
\begin{array}{l}
\dot{P} (x,t) +  \partial_x P(x,t)  = -u_{+} P(x,t) + u_{-}  Q(x,t)  + \Omega_{\Gamma}[P(x,t), Q(x,t)], \\ \\ 
\dot{Q} (x,t) -  \partial_x P(x,t)  = +u_{+} P(x,t) - u_{-} Q(x,t) - \Omega_{\Gamma}[P(x,t), Q(x,t)], \\ \\
\Omega_{\Gamma}[P(x,t), Q(x,t)]=  a P(x,t)  \int_{x}^{\infty}  {\cal G}\left[ y- \langle X(t) \rangle \right]  Q(y,t) dy   +\\  \\\quad\quad\qquad\qquad \qquad\qquad\qquad\qquad\qquad\qquad\qquad\qquad
 b Q(x,t) \int_{x}^{ \infty}  {\cal G} \left[y- \langle X(t) \rangle\right]P(y,t) dy,
\end{array}
\right.
\end{equation}}

\noindent with $u_{-}, u_{+}, a, b \in \mathbb{R}^{+}$ and initial conditions $P_0(x)$ and $Q_0(x)$. Associated with Eq.(\ref{TELEG}), we have  the normalisation constraint and the definition:

\begin{equation}
\label{NORM}
\begin{array}{l} 
\int_{\mathbb{R}} \left[P(x,t) + Q(x,t)\right] dx \equiv 1, \qquad \qquad\qquad{\rm (normalisation \,\, of \,\, the\,\, total\,\, probability \,\, mass) }\\ \\ 
\langle X(t) \rangle : = \int_{\mathbb{R}}  x\left[ P(x,t) + Q(x,t)\right] dx , \quad {\rm (barycenter \,\, location,\, of \,\, the\,\, total\,\, probability \,\, mass)}

\end{array}
\end{equation}

\vspace{0.3cm}

\noindent For the nonlinear dynamics Eq.(\ref{TELEG}), we will  now establish:

\vspace{0.3cm}
\noindent {\bf Proposition 3} 

\noindent  {\it For  the dynamics defined by Eq.(\ref{TELEG}) with   the choices $\omega \in [0,1]$,  $\eta \in ]-2, + \infty]$, the couple of  positive switching rates $u_{+}>0$, $u_{-}>0$   solving:

\begin{equation}
\label{RATIOS}
{u_{+}\over (1-\omega) }- {u_{-}\over(1+ \omega) }= 2+ \eta
\end{equation}

\noindent and  with the barycentre weight function:
  $${\cal G}(x) \equiv{\cal G}_{a, b, \eta} (x) = {(2+ \eta)\over 2(a+b) } B\left({1 \over2}, 1 + {\eta \over 2} \right) \cosh^{\eta}(x),$$

\noindent  Eq.(\ref{TELEG})  is solved by the soliton  waves:

\begin{equation}
\label{SOLVAR1}
{P(x- \omega t)  \over (1 + \omega)}= {Q(x-\omega t) \over (1-\omega)} = {1 \over2}B\left( {1 \over 2}, 1+ {\eta\over 2} \right) \left[\cosh(x- \omega t)\right]^{-(2+ \eta)}.
\end{equation}}


\vspace{1cm}
\noindent {\bf Proof of Proposition 3}

\noindent We introduce the change of variables $t \mapsto \tau$ and $x \mapsto \xi = (x - \omega t)$ and we  focus on  the  stationary regime  $\partial_{\tau} P(\xi, \tau)= \partial_{\tau}Q(\xi, \tau)=0$. We assume the symmetry  $P(\xi) = P(-\xi)$ and $Q(\xi) = Q(-\xi)$ and so  Eq.(\ref{SOLVAR1}) implies:

    $$\langle X(t) \rangle = \int_{\mathbb{R}}  (\xi+ \omega t) \left[ P(\xi) + Q(\xi)\right] d\xi=\omega t.$$

\noindent  In the stationary regime,    Eq.(\ref{TELEG})  can be rewritten as  as :

\footnotesize{\begin{equation}
\label{STAT}
(1- \omega) \partial_{\xi}P(\xi) = (1+ \omega) \partial_{\xi}Q(\xi) =  a \, P(\xi) \int_{\xi }^{\infty}  {\cal G}_{a,b, \eta}(y) Q(y) dy +  b\, Q(\xi) \int_{\xi }^{\infty}  {\cal G}_{a,b,\eta}(y) P(y) dy - u_{+} P(\xi) + u_{-}Q(\xi).
\end{equation}}

\noindent Introducing  the  rescaling factors:

\begin{equation}
\label{RESC}
\left\{
\begin{array}{l}
P(\xi) := (1+ \omega) \hat{P}(\xi) \qquad {\rm and }\qquad Q(\xi) := (1- \omega) \hat{Q}(\xi) , \\ \\ 
u_{+} := \hat{u}_{+}(1-\omega)\qquad  {\rm and} \qquad u_{-} := \hat{u}_{-} (1+\omega),
\end{array}
\right.
\end{equation}

\noindent  we can rewrite Eq.(\ref{STAT}) as:

\footnotesize{\begin{equation}
\label{RESCRW}
\partial_{\xi} \hat{P} (\xi) = \partial_{\xi} \hat{Q}(\xi) = - \hat{u}_{+} \hat{P}(\xi)  +  \hat{u}_{-} \hat{Q } (\xi) + a\,  \hat{P}(\xi) \int_{\xi }^{\infty}  {\cal G}(a,b,\eta)(y) \hat{Q}(y) dy + b\,  \hat{Q}(\xi) \int_{\xi }^{\infty}  {\cal G}(a,b,\eta)(y) \hat{P}(y) dy .
\end{equation}}
\noindent In view of Eqs.(\ref{SOLVAR1}) and (\ref{RESCRW}), we now assume that $ {\cal G}_{a,b, \eta}(\xi) =  {\cal A} (a,b,\eta) \left[ \cosh(\xi) \right]^{\eta}$ and  $\hat{P} (\xi) = \hat{Q}(\xi) = \hat{ {\cal N}}(\eta) \cosh^{-(2+\eta)}(\xi) $. and a direct substitution into  Eq.(\ref{RESCRW}) yields:

{\footnotesize\begin{equation}
\label{FF1}
\begin{array}{l}
-\hat{{\cal N} }(\eta)(2 + \eta)  \left[\cosh(\xi) \right]^{-(\eta +3)} \sinh(\xi) = \left[\hat{u}_{-} - \hat{u}_{+} \right]\hat{{\cal N} }(\eta)\left[\cosh(\xi) \right]^{-(2 + \eta)} + \\ \\
\qquad \qquad \qquad \qquad \qquad \qquad \qquad \qquad  \hat{ {\cal N}}^{2}(\eta) (a+b)  
\left[\cosh(\xi) \right]^{-(2 + \eta)} 
\underbrace {\int_{\xi }^{\infty} { {\cal A}(a,b, \eta) \over \left[ \cosh(\xi)\right]^{2} } d\xi}_{ {\cal A}(a,b, \eta)(1- \tanh(\xi)) }.
\end{array}
\end{equation}}

\noindent By direct identification, one concludes that one  has to fulfil:

\begin{equation}
\label{SOLVAR2}
\left\{
\begin{array}{l}
 2+\eta=  {\cal A}(a,b, \eta) (a+b) \hat{{\cal N}}(\eta) \quad \Rightarrow \quad {\cal A}(a,b, \eta) = {2+ \eta \over(a+b) \hat{{\cal N}}(\eta)}, \\ \\
 \left[\hat{u}_{-} - \hat{u}_{+} \right] +  {\cal A}_{a, b,\eta} (a+b)  \hat{{\cal N}}(\eta)=0 \quad \Rightarrow \quad   \left[\hat{u}_{+} - \hat{u}_{-} \right]  =2+\eta .\end{array}
\right.
\end{equation}

\noindent Finally, the first line in  Eq.(\ref{NORM}) implies:
$$
2\hat{{\cal N}}^{-1}(\eta) = \int_{\mathbb{R}}  (1 + \omega) {d\xi \over \left[ \cosh(\xi) \right]^{2 + \eta}} + \int_{\mathbb{R}}  (1 -\omega) {d\xi \over \left[ \cosh(\xi) \right]^{2 + \eta}} = B\left({1 \over2}, 1 + {\eta \over 2} \right).
$$

\rightline{{\bf End of the Proof }} 

\vspace{0.5cm}
\noindent {\bf Remark 3}. Here again we  observe that a normalised soliton cannot be generated  for modulation kernels ${\cal G}(z) \propto\cosh^{\eta} (z)$  for $\eta < -2$ and therefore, similarly to  section \ref{difon}, one   concludes that $\eta =2$  is the bifurcation threshold  separating two drastically  swarm propagation modes


\subsection{Corresponding mean-field game dynamics for piecewise deterministic evolutions}

\noindent Following the development  followed in section \ref{difon} where  Brownian motion environments drive the dynamics, we now  construct a MFG dynamics  with telegraphic noise driving which leads  to ergodic regimes similar to  the  soliton found in Proposition 3.  To this aim,  consider the controllable piecewise evolution dynamics:

\footnotesize{ \begin{equation}
\label{TELEGMFG}
\left\{
\begin{array}{l}
\partial_tP(x,t) +  \partial_x P(x,t)  = -u_{+}(x,t)  P(x,t) + u_{-}(x,t)  Q(x,t)   , \\ \\ 
\partial_tQ (x,t) -  \partial_x Q(x,t)  = +u_{+}(x,t)  P(x,t) - u_{-}(x,t)  Q(x,t), 
\end{array}
\right.
\end{equation}}

\noindent which differs from Eq.(\ref{TELEG}) by the fact  that the Poisson switching rates $u_{-}(x,t) $ and $u_{+}(x,t)$ are now explicitly  $(x, t)$-dependent. For a time horizon $t \in [0,T]$, let us introduce a couple of cost functions ${\cal J}_{\pm}$ in the form  discussed in [19]:

\begin{equation}
\label{COST}
\left\{
\begin{array}{l}
{\cal J}_{\pm}\left[ X(\cdot) , u_{\pm}(\cdot)\right] = \mathbb{E} \left\{ \int_{0}^{T} \left\{{\cal L} (u_{\pm }(X(s), s)  +{\cal W}[P(\cdot),s), Q(\cdot), X(s),s) ]ds\right\}  \right\} + C_{\pm, T}(X(T))\\ \\
{\cal L} (u_{\pm}(x,t),t) :=  u_{\pm}(x,t)  \ln  \left[u_{\pm}(x,t) \right]- u_{\pm}(x,t)  +1
\end{array}
\right.
\end{equation}

\noindent  where $C_T(X(T))$ stands for a  final cost. In Eq.(\ref{COST}), the  running cost  ${\cal W}[P(x,t),Q)(x,t)]$  depends only on the probability densities $P(x,t)$ and $Q(x,t)$. This functional structure confers to the dynamics  its MFG character.  The objective is now  to  minimise the global costs  ${\cal J}_{\pm}\left[ X(\cdot) , u_{\pm}(\cdot)\right] $ by optimally adjusting  the switching  rates $u_{\pm}(x,t)$. Invoking the dynamic programming (DP) principle, we may now derive the associated  Hamilton-Belman-Jacobi (HBJ)  equation and for the  resulting couple of value functions $V_{\pm}(x,t)$, we obtain:

\begin{equation}
\label{HBJ}
\left\{
\begin{array}{l}
\partial_tV_{+}(x,t) +  \partial_x V_{+} (x,t) + \min_{u_{+} } \left\{ {\cal L} (u_{+ }(x,t), t) + u_{+} \left[ V_{-} (x,t)- V_{+} (x,t)\right] \right\}  + {\cal W} (P(x,t),  Q(x,t) )=0 ,\\ \\ 
\partial_tV_{-}(x,t)  - \partial_x V_{-} (x,t)  + \min_{u_{-} } \left\{ {\cal L} (u_{- }(x,t), t) + u_{-} \left[ V_{+} (x,t)- V_{-} (x,t)\right] \right\}  + {\cal W}(P(x,t),  Q(x,t) ) =0.
\end{array}
\right.
\end{equation}

\noindent Performing the required minimisations, Eq.(\ref{HBJ}) becomes [19]:

\begin{equation}
\label{HBJ1}
\left\{
\begin{array}{l}
\partial_tV_{+}(x,t) +   \partial_x V_{+} (x,t) +\left[ 1- e^{+V_{+} (x,t) -V_{-} (x,t)}\right]  + {\cal W} (P(x,t),  Q(x,t) ) =0,\\ \\ 
\partial_tV_{-}(x,t)  - \partial_x V_{-} (x,t)  + \left[ 1- e^{+V_{-} (x,t) - V_{+} (x,t)}\right]  + {\cal W}(P(x,t),  Q(x,t) )=0
\end{array}
\right.
\end{equation}

\noindent and the  optimal switching rates $u^{*}_{+}(x,t)$ and $u^{*}_{-}(x,t)$ are given by:

\begin{equation}
\label{RATES}
\left\{
\begin{array}{l}
u^{*}_{+}(x,t) =  e^{+V_{+}(x,t) - V_{-}(x,t)},\\ \\ 
u^{*}_{-}(x,t) =e^{+V_{-}(x,t) - V_{+}(x,t)}.
\end{array}
\right.
\end{equation}

\vspace{0.5cm}
\noindent {\bf Proposition 4}.

\noindent {\it  Given   $\omega \in [0,1]$ and  with the cost in  Eq.(\ref{COST})  defined as:

\begin{equation}
\label{KERNEL }
\left\{
\begin{array}{l}
{\cal W}(P, Q) = g(q ,\omega) \left[PQ\right]^{q}, \qquad q >0,\\ \\
g(q ,\omega)= { (q+1)(1 - \omega^{2})^{(2q-2)} \over 4^{q} q^{2} ( 2 - \omega^{2})  }  \left[ B\left({1 \over 2} , {1 \over 2q}\right) \right]^{2q},
\end{array}
\right.
\end{equation}

\noindent   the probability densities $P(x,t)$ and $Q(x,t)$ solving the  ergodic regime of the MFG Eq.(\ref{HBJ1})  are given by  the soliton waves:

\begin{equation}
\label{ERGODICSOL}
\left\{ 
\begin{array}{l}
 (1- \omega)  P(x - \omega t)   = (1 + \omega)Q(x-\omega t)   = {\hat{{\cal N}}(q) \over \left[ \cosh (x - \omega t) \right]^{1/q}} , \\ \\
 
 \left[\hat{{\cal N}}(q)\right]^{-1} =  {(1- \omega^{2}) \over2}B\left({1 \over 2} , {1 \over 2q}\right).
\end{array}
\right.
\end{equation}}

\vspace{1cm}
\noindent {\bf Proof of Proposition 4}.

\noindent Let us  introduce  the following transformations\footnote{In the sequel, for  simplicity of the notation, we  shall  omit to repeat  the ubiquitous  $(x,t)$ argument.}:

\begin{equation}
\label{TRA1}
\left\{
\begin{array}{l}
\varphi_{A} = e^{-\epsilon t -V_{+} } \qquad {\rm and } \qquad  \Gamma_{A} = P e^{\epsilon t +V_{+}} \,\, \Rightarrow P= \Gamma_A \, \varphi_A ,\\ \\

\varphi_{B} = e^{-\epsilon t -V{-}} \qquad {\rm and } \qquad \Gamma_{B} = Q e^{ \epsilon t+V_{-}} \,\, \Rightarrow Q= \Gamma_B\,  \varphi_B,

\end{array}
\right.
\end{equation}

\noindent with $ \epsilon \in \mathbb{R}^{+}$.  From Eqs.(\ref{RATES}) and (\ref{TRA1}),  we have: 

\begin{equation}
\label{UU}
u^{*}_{+} = {\varphi_{B}  \over \varphi_{A} },\qquad {\rm and} \qquad u^{*}_{-} = {\varphi_{A}  \over \varphi_{B} }.
\end{equation}

\noindent  Using Eq.(\ref{TRA1}) and substituting the definitions of $\varphi_A$ and $\varphi_B$  into Eq.(\ref{HBJ1}), we obtain:

\begin{equation}
\label{TRA2}
\left\{
\begin{array}{l}
\partial_{t} [\varphi_{A}] + \partial_{x} [\varphi_{A}]  -  \varphi_{A} + \varphi_{B} -\left[ {\cal W} - \epsilon \right]\, \varphi_{A}=0,  \qquad \qquad i)  \\ \\
\partial_{t} [\varphi_{B}] - \partial_{x}[ \varphi_{B}]  + \varphi_{A} -  \varphi_{B} - \left[{\cal W}- \epsilon\right] \, \varphi_{B}=0.  \qquad \qquad ii) 
\end{array}
\right.
\end{equation}

\noindent Introducing  $u^{*}_{+}(x,t)$ and $u^{*}_{-}(x,t)$ given by Eq.(\ref{RATES}) into Eq.(\ref{TELEGMFG})  and using once more Eq.(\ref{TRA1}), we end with:

\begin{equation}
\label{TRA3}
\left\{
\begin{array}{l}
\partial_{t} [\Gamma_{A}] + \partial_{x} [\Gamma_{A}]  + \Gamma_{A} - \Gamma_{B} +\left[{\cal W}-\epsilon \right]\, \Gamma_{A}=0, \qquad \qquad iii) \\ \\ 

\partial_{t} [\Gamma_{B}] - \partial_{x}[ \Gamma_{B}]  - \Gamma_{A} + \Gamma_{B} +\left[  {\cal W} -\epsilon \right] \, \Gamma_{B}=0.  \qquad \qquad  iv)
\end{array}
\right.
\end{equation}

\noindent In view of  Eqs.(\ref{TRA2}) and (\ref{TRA3}) and the set of  definitions  introduced in Eq.(\ref{TRA1}),  we can derive:

\begin{equation}
\label{EQP}
\left\{
\begin{array}{l}

 \left[ i) \times \Gamma_A\right] +\left[  iii)  \times \varphi_A\right]\,\, \Rightarrow \,\,  \partial_t P  + \partial_x P + \left\{\Gamma_A \varphi_B  - \varphi_A \Gamma_B\right\}=0, \\ \\
\left[ ii) \times \Gamma_B\right] + \left[ iv)\times \varphi_B \right]\,\, \Rightarrow \,\, \partial_t Q  -\partial_x Q   - \left\{\Gamma_A \varphi_B  - \varphi_A \Gamma_B\right\}=0. 
\end{array}
\right.
\end{equation}

\noindent Performing the change of variables:

\begin{equation}
\label{CVAR}
\left\{
\begin{array}{l}
t \mapsto \tau, \quad x \mapsto \xi = (x-\omega t), \\ \\ 
\partial_ t \mapsto \partial_{\tau} - \omega \partial_{\xi}  \qquad \partial_t \mapsto \partial_{\tau}, \\ \\
 \hat{P}= :P(1-\omega), \quad \hat{Q}=: Q(1+\omega) ,
\end{array}
\right.
\end{equation}

 \noindent enables to rewrite Eq.(\ref{EQP}) as:

\begin{equation}
\label{EQPHAT}
\left\{
\begin{array}{l}
  {1  \over (1- \omega)}  \partial_{\tau} \hat{P}+   \partial_{\xi}\hat{P}+ \left\{\Gamma_A \varphi_B  - \varphi_A \Gamma_B\right\}=0, \\ \\
 {1  \over (1+ \omega)}  \partial_{\tau} \hat{Q}   -\partial_{\xi}\hat{Q }  - \left\{\Gamma_A \varphi_B  - \varphi_A \Gamma_B\right\}=0.
\end{array}
\right.
\end{equation}

\noindent Let us now focus  on stationary regimes  for which $\partial_{\tau}\hat{P} = \partial_{\tau} \hat{Q}=0$. Using  Eq.(\ref{EQPHAT}) and  since  normalisation imposes that 
$\lim_{|\xi |\rightarrow \infty} \hat{P}(\xi) =0 $ and  $\lim_{|\xi |\rightarrow \infty} \hat{Q}(\xi) =0 $,  we have $\hat{P}(\xi) = \hat{Q}(\xi) $ and therefore:

\begin{equation}
\label{AS}
\hat{P}(\xi) = (1 - \omega)P(\xi)  =  (1 - \omega) \varphi_A (\xi)\Gamma_A (\xi) = \hat{Q} (\xi)  =(1 + \omega) Q /\xi) = (
1 + \omega) \varphi_B (\xi)\Gamma_B(\xi) ,
\end{equation}

\vspace{0.3cm}
\noindent Since  $\hat{P}(\xi)  = \hat{Q}(\xi) $, the first  two lines in Eq.(\ref{EQPHAT}) imply that we can write :

\begin{equation}
\label{EQPDERIV}
\partial_{\xi \xi} \hat{P} + \partial_{\xi} \left\{\Gamma_A \varphi_B  - \varphi_A \Gamma_B\right\}=0.
\end{equation}

\noindent  In the stationary regime,  Eqs.(\ref{TRA2}) and (\ref{TRA3})  enable to write  straightforwardly the following combinaisons:

$$
\begin{array}{l}
\varphi_B \partial_{\xi} \Gamma_A = - { \varphi_B \Gamma_A\over (1- \omega)} + { \varphi_B \Gamma_B\over (1- \omega)} - { \varphi_B \Gamma_A ( {\cal W} - \epsilon) \over (1- \omega)}, \\  \\ 
\Gamma_A \partial_{\xi} \varphi_B = +{ \varphi_A \Gamma_A\over (1+\omega) }- { \varphi_B \Gamma_A\over (1+\omega) } - { \varphi_B \Gamma_A({\cal W} - \epsilon) \over (1+\omega) }, \\ \\
\Gamma_B\partial_{\xi}\varphi_A = + { \varphi_A \Gamma_B\over (1- \omega)} -  { \varphi_B \Gamma_B\over (1- \omega)} + {  \varphi_A \Gamma_B({\cal W} - \epsilon) \over (1-\omega) }, \\ \\
\varphi_A \partial_{\xi} \Gamma_B = -{ \varphi_A \Gamma_A\over (1+\omega) }+ { \varphi_A \Gamma_B\over (1+\omega) } +{ \varphi_A \Gamma_B({\cal W} - \epsilon) \over (1+\omega) }.
\end{array}
$$

\noindent

\noindent This enables to write :

\begin{equation}
\label{CROSS2}
\partial_x \left( \Gamma_B \varphi_A - \Gamma_A \varphi_B\right)=  -{2 \over (1- \omega^{2})} \left\{
(\hat{P}+\hat{Q}) +\left(  \epsilon - 1 -{\cal W}\right) \left( \varphi_A \Gamma_B +  \varphi_B \Gamma_A\right) \right\} .
\end{equation}

\noindent Consistent with Eqs.(\ref{TRA1}) and (\ref{AS}), we now assume that:

\begin{equation}
\label{HYP}
\varphi_B = \varphi_A(1- \omega) \qquad {\rm and } \qquad \Gamma_B = \Gamma_A {1 \over (1 + \omega) } \quad \Rightarrow \quad   \left(\varphi_A \Gamma_B +  \varphi_B \Gamma_A\right)=  \left({2- \omega^{2}  \over1- \omega^{2}}\right) \hat{P}.
\end{equation}

\noindent Since  $\hat{P} (\xi) = \hat{Q}(\xi)$,  Eqs.(\ref{EQPDERIV}), (\ref{CROSS2}) and (\ref{HYP}) imply:

\begin{equation}
\label{SCH}
\begin{array}{l} 
\partial_{\xi \xi} \hat{P}(\xi)  = { 2 \over (1- \omega^{2})} \left [2  + \left(  \epsilon -1 -{\cal W}(\hat{P}, \hat{Q}) \right)  \left({2- \omega^{2}  \over1- \omega^{2}}\right)\right] \hat{P}(\xi) 
\end{array}
\end{equation}

\noindent   We now focus on the set of  running cost functions,:

\begin{equation}
\label{PATATOS}
{\cal W} (P, Q) =  {\cal W}(PQ) = {\cal W}( {\hat{P} \hat{Q} \over 1-\omega^{2}})   : = g\left[ \hat{P}(\xi)\right]^{2 q}, \qquad  g, q \in \mathbb{R}^{+}.
\end{equation}

\noindent It is immediate to realise  that Eq.(\ref{SCH})  exhibits  the standard form of the nonlinear Schr\"odinger equation:

 \begin{equation}
\label{SCH2}
 \partial_{\xi \xi} \hat{P}(\xi)= 2{  [ \epsilon(2  -  \omega^{2}) -\omega^{2}] \over (1 - \omega^{2} )^{2}} \hat{P}(\xi) - g {  (2- \omega^{2}) \over (1 - \omega^{2} )^{2}}\left[ \hat{P}(\xi)\right]^{2q+1}.
\end{equation}

 \noindent Provided appropriate constants $g, q,\epsilon, \omega$ are chosen,Eq.(\ref{SCH2}) can be integrated to yield a soliton which coincides with  the one found   in Eq.(\ref{SOLVAR1}). To see this, multiply both sides of Eq.(\ref{SCH}) by $\partial_{\xi}P$ and integrate once with respect to $\xi$ (with zero integration constant), we obtain:

 \begin{equation}
\label{NLSCH}
 \left( \partial_\xi \hat{P}(\xi)\right)^{2} = 2  {  [ \epsilon(2 - \omega^{2}) -2] \over (1 - \omega^{2} )^{2}} \left[\hat{P}(\xi)  \right]^{2}  - g {  (2- \omega^{2}) \over  (q+1)(1 - \omega^{2} )^{2}} \left[\hat{P}(\xi)\right]^{2q+2}= 0.
\end{equation}

\noindent Using  once again the separation of variable technique, Eq.(\ref{NLSCH})  leads to:

\begin{equation}
\label{SEP}
\left\{
\begin{array}{l}
\int^{\xi}  {d\hat{P}(\xi) \over \sqrt{A_1(\epsilon, \omega)   \hat{P}^{2}(\xi) -  A_2(g, \omega, q)\left[\hat{P} (\xi) \right]^{2q+2} }}= \xi, \\ \\ 
A_1(\epsilon, \omega):= 2{  [ \epsilon(2  -  \omega^{2}) -\omega^{2}]\over (1 - \omega^{2} )^{2}}, \\ \\ 
A_2(g, \omega,q):= g {  (2- \omega^{2}) \over (q+1)(1 - \omega^{2} )^{2}}.
\end{array} 
\right.
\end{equation}

\noindent   Now we can verify that the {\it Ansatz}  $\hat{P} (\xi)= { \hat{{\cal N}}(q)  \over \left[ \cosh(\xi)\right]^{1/q} }$ with $q>0$ solves Eq.(\ref{SEP}) provided we impose :

\begin{equation}
\label{FINALSOLIT}
\left\{
\begin{array}{l}
1 =  A_1(\epsilon, \omega)  q^{2} \quad \Rightarrow\quad \epsilon =  {1 \over (2- \omega^{2})}\left[ {(1-\omega^{2})^{2} \over 2q^{2}}  + \omega^{2}\right]\\ \\

1= A_2(g, \omega,q) q^{2}\left[ \hat{{\cal N}}(q)\right]^{2q}.
\end{array}
\right.
\end{equation}
 
 \noindent So given the couple constants $g>0$ and $\omega \in [0,1]$, we choose $g$ to satisfy the last line of Eq.(\ref{FINALSOLIT}) and then  $ \epsilon$  follows. Since $\hat{P}(\xi)= \hat{Q}(\xi)$,  the normalisation factor $\hat{{\cal N}}(q) $ is given by:
 
 $$
 \int_{\mathbb{R}} \left[P(\xi)  +(Q (\xi)  \right] d\xi = 1  \, \, \ \Rightarrow \, \, \ \left[ \hat{{\cal N}}(q) \right]^{-1}
= {2 \over (1- \omega^{2})} \int_{\mathbb{R}}{d \xi \over\left[ \cosh(\xi) \right]^{1/q}} = {2 \over (1- \omega^{2})} B\left({1 \over 2} , {1 \over 2q}\right).
 $$
 
 \rightline{{\bf End of the proof}}

 \vspace{1cm}
\noindent  {\bf Remark 4}. Fixing  $\omega$ as  given  by Eq.(\ref{RATIOS}) and choosing  $1/q = (2 + \eta)>0$,  once more  one  observes that the  solitons given in  Eqs.(\ref{SOLVAR1}) and (\ref{ERGODICSOL})  coincide.  Hence under telegraphic noise environments,  the soliton generated by  exogenous interaction rule Eq. (\ref{DIFFNL}) coincide, in the ergodic regime, with the optimal solution of a MFG with an  appropriate choice of the running cost function.

\vspace{1cm}
\noindent  {\bf  Remark 5}. As shown in [17] (see Chapter 9),  the WGN can be derived from the telegrapher's process by an ad-hoc rescaling of the  switching rates and the jumps sizes. Exploiting this observation, several contributions [20, 21, 22] show how the parabolic  Burgers' equation describing  the diffusive dynamics  of section \ref{difon} coincides, via an  an ad-hoc limiting procedure, with  the hyperbolic discrete velocity Boltzmann equation which describes  the piecewise deterministic  dynamics used  of section \ref{trifon}. In other words the RW dynamics [17]  is a generalisation of the Burgers's equation.  Along the same lines,  the assertions made in our present   Propositions  3 and  4 are  generalisations of the assertions made in Propositions 1 and 2.

 \section{Conclusion}
 
 \noindent To a stationary collective motions sustained by  exogenously defined interactions' rules, it  corresponds  a mean-field games (MFG)  with optimal stationary equilibria yielding the  same collective evolution. This parallel  is analytically exemplified here in situations where the driving  stochastic  environment  is either a Brownian motion or a  two-states Markov chains in continuous time. For  long-range interacting   scalar agents each evolving on the  real line,  we explicitly show  the existence of  bifurcation threshold separating two drastically  different swarm propagation modes: one either observes a stable soliton  or a diffusive evanescent wave. As in the Kuramto's dynamics for agents evolving on a compact state space, the  transition is due to the competition between a synchronisation effect due to the mutual  interactions and the  desynchronisation due to the random environment. Since our proposed models  are  exactly solvable,  they hopefully  enrich the yet rather scarce  collection  of fully solvable models relevant for  multi-agents dynamics.

 \section*{Appendix}
 
 \noindent This goal of this appendix is to derive the second line in Eq. (\ref{FORBACK2}). Using the fact that $\rho = \Phi \, \Psi$ and $ u= -\mu \sigma^{2} \ln \Phi - \epsilon t$, the FPE  Eq.(\ref{FORBACK}) reads\footnote{All arguments  of the scalar fields $\Phi(x,t)$ and $\Psi(x,t)$ are omitted on this Appendix.}:
 
 $$
( \partial_t \Phi) \Psi +  \Phi (\partial_t \Psi)  = 
\partial_x \left\{ {1 \over \mu }   \left[ - \mu \sigma^{2} { \partial_x \Phi \over \Phi }\right]\Phi \Psi   - b \Phi \Psi \right\}  + {\sigma^{2} \over 2} \Psi \partial_{xx} \Phi + \sigma^{2} (\partial_x \Phi )(\partial_x \Psi) + {\sigma^{2} \over 2} \Phi\partial_{xx} \Psi 
 $$

 \noindent or equivalently:

 $$
( \partial_t \Phi) \Psi +  \Phi (\partial_t \Psi)  = 
\partial_x \left\{ - b \Phi \Psi \right\}  - {\sigma^{2} \over 2} \Psi \partial_{xx} \Phi.
+ {\sigma^{2} \over 2} \Phi\partial_{xx} \Psi 
 $$

 \noindent This can be rewritten as:
 
 \begin{equation}
\label{INTERM}
 \Psi \left\{ \partial_t \Phi + b \partial_x \Phi +{ \sigma^{2} \over 2}  \partial_{xx} \Phi \right\} = - \Phi\left\{  \partial_t \Psi + b \partial_x \Psi -{ \sigma^{2} \over 2}  \partial_{xx} \Psi \right\}.
\end{equation}

 \noindent The first line in Eq.(\ref{FORBACK2})  implies:

 $$
\left\{ \partial_t \Phi + b \partial_x \Phi +{ \sigma^{2} \over 2}  \partial_{xx} \Phi \right\}  =  {1 \over \mu \sigma^{2}} \left\{  \epsilon\Phi - g \left[ \Phi \Psi\right]^{a}   \Phi\right\} .
 $$

 \noindent Hence Eq.(\ref{INTERM}) takes the form:

 $$
  \Psi {1 \over \mu \sigma^{2}} \left\{  \epsilon\Phi  - g \left[ \Phi \Psi\right]^{a}   \Phi\right\} = - \Phi\left\{  \partial_t \Psi + b \partial_x \Psi -{ \sigma^{2} \over 2}  \partial_{xx} \Psi \right\}.
 $$

 \noindent or dividing by $-{\Phi \over \mu \sigma^{2}}$, we obtain:

 $$
- \epsilon \Psi + g \left[ \Phi \Psi\right]^{a}   \Psi=  \mu \sigma^{2} \left\{ \partial_t \Psi + b \partial_x \Psi -{ \sigma^{2} \over 2}  \partial_{xx} \Psi   \right\}
 $$
 
 \noindent and hence  the second line in Eq.(\ref{FORBACK2}) follows:
 
 $$
\epsilon \Psi  + \mu \sigma^{2}  \partial_t \Psi=  -\mu \sigma^{2}  b \partial_x \Psi  +  {\mu  \sigma^{4} \over 2}  \partial_{xx} \Psi  + g \left[ \Phi \Psi\right]^{a}   \Psi.
 $$
 
\section{References}
\vspace{0.3cm} 
\noindent [1]  J. A. Acebr\'on, L. L. Bonilla, J. P\'erez Vicente, F. Riort and R. Spiegler. {\it The Kuramoto Modeé: A Simple Paradigm for Synchronization Phenomena}. Rev.  Mod. Phys. {\bf 77}, 137-185, (2005).

\vspace{0.3cm} 
\noindent [2] H. Yin. P. G. Mehta, S. P. Meyn and U. V. Shanbhag. {\it Synchronisation of Coupled Oscillators is a Game}.  IEEE Trans. Automatic Control {\bf 57}(4), 920-935, (2012).

\vspace{0.3cm} 
\noindent [3] J.-M. Lasry and P.-L. Lions. {\it Mean Field Games}. Japan J. Math {\bf 2}, 229-260, (2007).

\vspace{0.3cm} 
\noindent [4] J.-M. Lasry and P.-L. Lions. {\it Jeux \`a Champs Moyen - Cas Stationnaire}. C. R. Math, Acad. Sci.  {\bf 343}, 619-625, (2006).

\vspace{0.3cm} 
\noindent [5] J.-M. Lasry and P.-L. Lions. {\it Jeux \`a Champs Moyen - Horizon Fini et Contr\^ole Optimal}. C. R. Math, Acad. Sci.  {\bf 343},679-684, (2006).

\vspace{0.3cm} 
\noindent [6]  M. Huang, P. E. Caines and R. P. Malham\'e. {\it Large Population Cost-coupled LQG Problems with Uniform Agents: Individual-mass Behaviour and Decentralised $\epsilon$-Nash Equilibria}. Commun.  Inf. Sys.  {\bf 6}, 221-252, (2006).

\vspace{0.3cm} 
\noindent [7] O. Gu\'eant, J.-M. Lasry  and P. L. Lions. {\it Mean Field Games and Applications}. Paris-Princeton Lectures on Mathematical Finance {\bf 2003} (2010).

\vspace{0.3cm} 
\noindent [8] R. Carmona and F. Delarue. {\it\underline{Probabilistic theory meanfield games with applications I \& II}.} Springer (2018).

\vspace{0.3cm} 
\noindent [9] A. Bensoussan, J.  Frehse and Ph Yam. {\it\underline{ Mean field games and mean field type control theory}. } Springer (2013).

\vspace{0.3cm} 
\noindent [10] D. Gomes, D. Pimentel and V. Voskanyan. {\it\underline{ Regularity theory for mean-field game systems }.} Springer (2016).

\vspace{0.3cm} 
\noindent [11] V. Kolokoltsov, J. Li and W. Yang. {\it Mean Field Games and Nonlinear Markov Processes.} rXiv:1112.3744, (2011).

\vspace{0.3cm}  
\noindent [12]  V. Kolokoltsov and O. A. Malafeyev. {\it\underline{ Many agent games in socio-economic systems:}} 

\noindent {\it \underline{corruption, inspection,  coalition building, network growth, security}.} Springer (2019).

\vspace{0.3cm} 
\noindent [13] C. Gardiner. {\it \underline{Stochastic Methods}}.  Springer Series in Synergetics, (fourth edition),  (2010).

\vspace{0.3cm} 
\noindent [14]  I. S. Gradshteyn and M. Ryzhik. {\it \underline{Tables of Integrals, Series and Products}}.

\vspace{0.3cm} 
\noindent [15] I. Swieciki, T. Gobron and D. Ullmo. {\it Schr\"odinger approach to mean-field games}. Phys Rev. Lett.  {\bf 116}, 128701, (2016).

\vspace{0.3cm} 
\noindent [16] P. Cardaliaguet, J.-M.  Lasry, P. Lions and  A. Poretta. {\it Long time average of mean field games, with a nonlocal coupling}. SIAM J. Contr. and Optim. {\bf 51}, 3558-3591, (2013).

\vspace{0.3cm} 
\noindent [17] W. Horsthemke and R. Lefever. {\it \underline{Noise Induced Phase Transitions:} } 

\noindent {\it \underline{Theory and Applications in Physics, Chemistry and Biology"}}. Springer Series in Synergetics, (1984), (see in particular Chapter 9).

\vspace{0.3cm} 
\noindent  [18]  T. W. Ruijgrok and T. T. Wu. {\it A Completely Solvable  Model of the Nonlinear Boltzmann Equation}. Physica A {\bf 113}, 401-416, (1982).

\vspace{0.3cm} 
\noindent [19] M.-O. Hongler,  M. Soner and L. Streit, {\it Stochastic Control for a Class of Random Evolution Models}. Appl. Math. Optim. {\bf 49}, 113-121, (2004).

\vspace{0.3cm} 
\noindent [20] M.-O. Hongler and  L. Streit, {\it A probabilistic connection between the Burgers and a discrete Boltzmann
equation}. Europhys. Lett. {\bf 12(}3), 193?197, (1990).

\vspace{0.3cm} 
\noindent [21] R. Filliger,  M. Soner and L. Streit, {\it Connection between an Exactly Solvable Stochastic Optimal Control Problem and a Nonlinear Reaction-Diffusion Equation} 
J. Oprim. Thory Appl.  {\bf 137}(3), 497-505,  (2008).

\vspace{0.3cm} 
\noindent [22] E. Gabetta and B. Perthame, {\it Probabilistic investigations on the explosion of solutions of the Kac equation with infinite energy initial distribution}. J. Appl. Probab. {\bf 45}(1), 95-106, (2008).

\end{document}